# Ammonia or methanol would enable subsurface liquid water in the Martian South Pole

Antifreeze and liquid water on Mars

Isabel Egea-González[1*], Christopher P. McKay[2], John E. Hallsworth[3], Alberto Jiménez-Díaz[4], Javier Ruiz[5]

[1]Departamento de Física Aplicada. Escuela Superior de Ingeniería. Universidad de Cádiz, 11519 Puerto Real, Cádiz, Spain.

[2]Space Science Division, NASA Ames Research Center, Moffett Field, California, 94035, USA.

[3]Institute for Global Food Security, School of Biological Sciences, Queen'sUniversity Belfast, 19 ChlorineGardens, Belfast, BT9 5DL, UK.

[4]Departamento de Biología y Geología, Física y Química Inorgánica, ESCET, Universidad Rey Juan Carlos, 28933, Móstoles, Madrid, Spain.

[5]Departamento de Geodinámica, Estratigrafía y Paleontología; Facultad de Ciencias Geológicas, Universidad Complutense de Madrid, 28040 Madrid, Spain.

* Corresponding author:isabel.egea@uca.es (Isabel Egea-González)

## Abstract

The notion of liquid water beneath the ice layer at the south polar layered deposits of Mars is an interesting possibility given the implications for astrobiology, and possible human habitation. A body of liquid water located at a depth of 1.5 km has been inferred from radar data in the South Polar Cap. However, the high temperatures that would facilitate the existence of liquid water or brine at that depth are not consistent with estimations of heat flow that are based on the lithosphere's flexure. Attempts to reconcile both issues have been inconclusive or otherwise unsuccessful. Here, we analyse the possible role(s) of subsurface ammonia and/or methanol in maintaining water in a liquid state at subsurface temperatures that are compatible with the lithosphere strength. Our results indicate that the presence of these compounds at the base of the south polar layered deposits can reconcile the existence of liquid water with previous estimations of surface heat flow.

## MAIN TEXT

### INTRODUCTION

Radar data obtained from the radar subsurface sounder that is aboard the Mars Express spacecraft that orbits Mars reveals bright subsurface reflections in the Planum Australe region of the planet. These reflections cover an area that is about 20-km wide that is located at 193° E, 81° S (Fig. 1). The data are interpreted as an extant body of liquid water located beneath the south polar layered deposits (SPLD) at a depth of 1.5 km (1). The existence of liquid water on Mars is particularly interesting because it might facilitate the existence of life (microbial contaminants released during space exploration and/or Martian life). Even at temperatures too low to allow the function of terrestrial microbes, aqueous milieu can apparently preserve cells in a viable condition over vast periods, likely hundreds of millions of years (2). Furthermore, bodies of liquid water would be valuable as a resource to enable human habitation (3).

However, the inference of liquid water, based on the interpretation of the bright reflections, is controversial because this liquid state requires temperatures that imply local surface heat flows higher than those predicted from lithosphere strength or those predicted using thermal history models (4). There are alternative interpretations to liquid water that explain the bright reflections: electrically conductive minerals (5), constructive interference of $CO_2$ layers (6) and hydrated clays (7), but each of these explanations has been disputed (8–10).

Some authors have suggested the presence of salts that depress the freezing point of water to reconcile the evidence suggesting the existence of liquid water with the estimations of the surface heat flow. Perchlorates have been favoured as the most-likely salts because they are widespread on Mars, including the North Polar Cap (11). However, surface heat flows calculated assuming the existence of perchlorates within the SPLD are still higher than expected (4). It has been argued that the eutectic point of the SPLD ice can be low enough to provide the expected heat flow if diverse perchlorate salts are concomitantly present in the ice layer (9). In this case, the presence of magnesium perchlorate and sodium perchlorate in water lowers the freezing point temperature to 180 K(12). However, the amount of perchlorates needed to reduce the freezing point to values compatible with the predicted surface heat flow is high for binary and ternary solutions (one or two salts in water): ~20 to 50% by mass (9,12).

There are reasons to think that the *in-situ* concentration of perchlorates at the SLPD is relatively low at depth. Firstly, the source for the perchlorates (from chlorine) is thought to be linked to surface radiation and/or atmospheric processes. Secondly, perchlorates are not found at high levels in Mars meteorites (e.g., SNCs) that sample deeper materials (< ppm levels), and, thirdly, a concentration of perchlorates of 20 to 50% by mass would involve an excess of mass at the SPLD (13). Furthermore, even if the required proportion of perchlorates exists, the thermal analysis of the region where the putative liquid water is located indicates that the overlying ice layer should be extremely porous in order to allow the basal melting of a mixture of perchlorates and water ice at the heat flows predicted from lithospheric strength (14). The thermal conductivity of any dust that exists within the perchlorate-containing ice layer must be identical to the thermal conductivity of the regolith along the entire layer (which has a thickness of 1.5 km) for the surface heat flow, derived from the lithospheric strength, to be consistent with the presence of liquid water (8). This is difficult to reconcile with the high density of the SPLD and with ice compaction at depth.

Another option to explain the origin of the putative liquid water is to consider that it was formed in the past, when the planet was warmer, and remained liquid since that time in a metastable state. However, it is not clear if supercooled perchlorate–$H_2O$ solutions are stable over geologic time scales (13).Thus, the presence of perchlorates within the south polar layered deposits cannot reconcile the existence of liquid water with the heat flow estimations.Hence, there is an ongoing conflict between evidence for liquid under the SPLD, the thermal requirements to facilitate and stabilise this liquid state, and the thermal properties of the layer above the putative liquid body.

Here, we explore the possible presence of non-salt antifreeze substance(s) in the putative liquid body, permitting it to persist in liquid state at temperatures as low as 170 to 180 K (-103 to -93°C). We show that these low antifreeze substances, ammonia and methanol, would eliminate the conflict between this liquid state and heat flow estimates. We also discuss the implications of the existence of ammonia and methanol for habitability of the liquid water body or lake under the SPLD.

**RESULTS**

**The background heat flow at the South Polar Region**

Previous studies have used the relation between the temperature profile and the mechanical strength of the lithosphere to calculate the upper limits to the present-day heat flow of Mars from the effective elastic thickness of the lithosphere. Because evidence of flexure in this region due to polar cap loading are unclear (15,16), the surface heat flows proposed following this approach are upper limits, ranging between 23 and 32 mW m$^{-2}$ (15,17). These values are much lower than the lower limit of ~70 mW m$^{-2}$ that was deduced in previous work that consider perchlorates as antifreeze for maintaining a liquid body within the polar cap (4). Even with very favourable conditions it is very difficult to obtain a heat flow lower than 40 mW m$^{-2}$ being compatible with a liquid layer (14).

For comparison purposes, we have re-evaluated the background (i.e., regional) heat flow at the South Polar region from the effective elastic thickness of the lithosphere, abbreviated as $T_e$ (see the Methods Section). Due to insufficient evidence of flexure, deriving an estimated value for $T_e$ in this region is not easily achievable. We have calculated an upper limit for the surface heat flow at the South Polar region by using a minimum value of $T_e$ = 150 km (15) (see Methods). We have also calculated heat flows for $T_e$ values in the range 255 to 360 km, which are the best fit $T_e$ values according to ref. 15 and ref. 18. We summarize our results in Table 1. The obtained upper limit for the heat flow is 27 mW m$^{-2}$, although for $T_e$ best fit values the heat flow would be between 19 and 22 mW m$^{-2}$. We therefore consider a range between 19 and 27 mW m$^{-2}$ for the background heat flow in the South Polar region of Mars. This range is consistent with the heat flow upper limit of 23.5 mW m$^{-2}$ proposed in previous work (15).

**Subsurface temperatures at the SPLD**

To identify the range of temperatures at the base of the SPLD that are compatible with the surface heat flow calculated from flexural models, we have calculated the surface heat flows ($F_s$) that correspond with basal temperatures between 165 and 190 K (-103 and -83 ºC). The surface heat flow ($F_s$) obtained from this basal temperature range is shown in Fig. 2.

Fig. 2 indicates that the low surface heat flows necessary to account for the limited flexion of the lithosphere under the South Polar Cap requires a basal temperature of well below 180 K (i.e., less than -93 ºC), which is much lower than the melting points of mixtures formed by perchlorates and water ice. Thus, the apparent presence of liquid water under the SPLD indicates the likely presence of one or more substances that act as antifreeze, such as ammonia or methanol.

The presence of ammonia reduces the solidus of the ammonia–water system to 176 K (-97 ºC) even when the concentration of ammonia is low. Table 2 is a summary of possible candidates to reduce the SPLD ice-layer freezing point. If we assume that the ammonia is located at the base of the ice layer, the surface heat flow required for melting is of 31.2 mW m$^{-2}$, which can be easily compatible with a surface heat flow of 19 to 27 mWm$^{-2}$ if we take into account some porosity in the ice layer. In the case of a mixture of methanol and water ice, the melting point decreases to 171 K (-102 ºC) for a wide range of methanol concentrations. The heat equation resolution provides a surface heat flow of 20.3 mW m$^{-2}$ for a basal temperature of 171 K. These results for ammonia and methanol are consistent with the low surface heat flow that is predicted using flexural models for this region, and may indicate the presence of compounds under the SPLD that are not usually considered at the surface of Mars.

**DISCUSSION**

**Ammonia and methanol as antifreeze on Mars**

The high effective elastic thickness of the lithosphere under the South Polar Cap is consistent with a cold lithosphere with upper surface heat flows limit ranging between 19 and 27 mW m$^{-2}$. The temperature at the base of the SPLD should be much lower than 180 K (-93 ºC) to be compatible with these surface heat flow estimations. Therefore, the presence of substances that greatly reduce the melting point of water ice is required to explain the plausibility of a liquid water layer. In particular, ammonia- and methanol-rich ices have melting points that are consistent with the predicted surface heat flows.

The notion of ammonia-water mixtures on Mars is one that has not been previously considered. There have been no definitive identifications of ammonia on the surface and atmosphere of Mars, which could be because ultraviolet radiation converts atmospheric ammonia to $N_2$ (19). Derivatised ammonia has been detected on sand from Bagnold Dunes through the Sample Analysis at Mars (SAM) instrument, but it is not clear if they have an endogenic origin (20). However, we do know that ammonia plays an important role in the presence of liquid water in many icy bodies and is abundant in the Solar System. There are also suggestions of ammonia as a primordial molecule on Mars (19). The analysis of Martian meteorites indicates that ammonia has been preserved in the subsurface since the Noachian period (21), where it has been protected from ultraviolet radiation and cosmic rays. Furthermore, there may be ongoing subsurface sources (biotic and abiotic) for ammonia (22, 23). Thus, the hypothesis of the presence of ammonia is not implausible. In this sense, if there are ammonia seeps under the ice, then the ice slab could protect it from ultraviolet radiation and ammonia could accumulate. In the case of leaks from the subsurface into the atmosphere, ammonia would be quickly destroyed and beessentially undetectable. Moreover, the presence of ammonia would be a source of isotopically primordial N and could help explain the $^{14}N/^{15}N$ in the Martian atmosphere (24). Methanol has been proposed as a primordial molecule on Titan (25) and it could be also present in the early Martian environment (26). Furthermore, methanol could be produced on Mars from methane (e.g., ref. 27). Methane has been observed in small abundances in the Martian atmosphere (28), and given the very short times of residence it must be supplied from the planetary interior (29). Maybe, the supply of methane needed for the potential production of methanol could occur directly below the surface and even in the liquid body. As in the case of ammonia, the low proportion of methanol that is required to reduce the melting point together with the short photochemical lifetime in the Martian atmosphere (74 days (27)) could complicate its detection in case of leaks to the atmosphere.

**Implications and perspectives for habitability**

The results presented in the current study indicate that the existence of liquid water below the South Polar Cap at the temperatures predicted from the lithosphere strength could suggest the presence of compounds under the ice slab that are rarely considered on Mars. Whereas data from geochemical analyses of Mars surface materials do not indicate the presence of methanol or ammonia, their presence in the subsurface of the SPLD is plausible. These compounds reduce the melting point of the ice to values of 171 and 176 K. These temperatures lead to surface heat flow of 31.2 and 20.3 mW m$^{-2}$ for ammonia- and methanol-rich ice, respectively. The obtained surface heat flows agree with the range of 19 to 27 mW m$^{-2}$ that is obtained from the limited flexion of the lithosphere.

Ammonia and some ammonium salts entropically destabilise the cell membranes and biomacromolecules of terrestrial microbes; a property known as chaotropicity (30–32). In contrast, methanol is known to be biophilic because in this regard its behaviour is close to that of water so it is relatively neutral (31, 33). This said, chaotropic substances at subzero temperatures that rigidify biomacromolecular structures can facilitate the flexibility of cellular macromolecules and so facilitate, rather than inhibit, the biotic activities of microbial systems (33, 34).

Although some terrestrial microbes are known to produce methane, a mechanism to explain the origin of these compounds under the SPLD is yet to be determined. Nevertheless, the perfect match between the thermal characteristics of these systems and the predicted thermal state of the South Polar Cap seems remarkable and should be considered as a possible solution to the thermal problem for the presence of liquid water under the south polar layered deposits.

## METHODS

### Surface heat flow from the effective elastic thickness of the lithosphere

We have employed the methodology outlined by ref. 17 and ref.35 to evaluate the regional heat flow at the South Pole using the effective elastic thickness of the lithosphere ($T_e$). $T_e$ is not easy to be estimated for this region because unclear evidence of flexure. The minimum acceptable value for $T_e$ would be 150 km in accordance with ref. 15, which is comparable to (or lower than) the minimum $T_e$ values obtained by other authors (18, 36, 37). Thus, because the inverse relation between lithosphere strength and heat flow, we use $T_e$= 150 km in order to derive an upper limit for the surface heat flow in the South Polar region. For comparison, we calculated heat flows for best-fit $T_e$ values in the range between 255 km (18) and 360 km (15), whereas flexure indetermination does not permit to obtain appropriate heat flow lower limits for this region.

We assume a mean crustal thickness of 70 km for the South Polar region, based on the updated crustal thickness model in ref.38. Because the thickness of the crust is much lower than the effective elastic thickness of the lithosphere, the crust must be mechanically welded to the mantle lithosphere. Because this and the absence of clear lithosphere flexure, is sufficient to consider the mechanical properties of the mantle lithosphere in the calculations. Thus, we consider zero plate curvature, rheological parameters for olivine(39), and a strain rate of $10^{-14}$ s$^{-1}$ appropriate for the temporal scale of polar cap loading(16).

The radioactive crustal heat production is taken to be the mean estimated for Mars from Mars Odyssey GRS measurements (40) ($4.89 \times 10^{-11}$ W kg$^{-1}$), whereas for the mantle lithosphere we assume 0.1 times the crustal heat production value (41). Densities of 2900 and 3500 kg m$^{-3}$ are used, respectively, for the crust and the mantle lithosphere. The polar cap thickness is taken as 3 km(36), and it is assumed to be composed of 85% water ice and 15% basaltic dust (15, 37, 42).

For the thermal conductivity of the crust we consider two possibilities, 2 and 2.5 W m$^{-1}$ K$^{-1}$; the first value would be appropriate for a basaltic crust, whereas the later could be considered an upper limit if the crust includes notable amounts of granite-type materials (see ref.43 for a review). For the mantle lithosphere, we use a temperature-dependent thermal conductivity (35). The thermal conductivity of the polar cap is estimated from a geometric mean (14) between those of water ice (which is temperature-dependent (44)) and basalt (the ice layer properties are described in detail in the next section). Finally, the surface temperature is taken as 155 K (-118 ºC) (37); note that for the calculation of regional heat flow we consider values representative for the region loaded by the polar cap load, not for the lake position. For further discussion of lithosphere parameters see ref.35.

## Surface heat flow and temperature at the base of the SPLD

To enable an analysis of the dependence of surface heat flows on the temperature at the depth of the putative liquid water, we have solved the steady state heat conduction equation for the overlying ice layer. The heat equation resolution requires to specify the thermal properties and the boundary conditions in the studied area.

The putative liquid water is located at a depth of 1.5 km under the SPLD, which are thought to be formed by a mixture of dust and water ice with a 1-m-thick layer of $CO_2$ ice on top (45). The thermal effect of the thin layer of $CO_2$ ice is negligible in a layer with a total thickness of 1.5 km, so we did not include $CO_2$ ice in our calculations. The density and the thermal conductivity of the layer formed by dust and water ice are required parameters to solve the heat equation. We assume a density of 1271 kg m$^{-3}$ (according to ref.37), which is consistent with a dust proportion of 15% by volume.

Regarding the thermal conductivity of the layer, we use a temperature-dependent thermal conductivity for the water ice given by the equation (44): $k_{ice}$ = 1.16 [1.91 − 8.66 · 10$^{-3}$$T$ + 2.97 · 10$^{-5}$$T^2$], where $k_{ice}$ is the water ice conductivity in W m$^{-1}$ K$^{-1}$, and $T$ is the temperature of the ice expressed in ºC. For the dust component, we assume a thermal conductivity for basalt rocks of 2 W m$^{-1}$ K$^{-1}$ (46), and we obtain the thermal conductivity of the mixture of water ice and dust as a geometric mean of both components.

The surface and basal temperatures are the boundary conditions for the heat conduction equation. The annual average surface temperature at the location of the putative subsurface liquid body is constrained to be ~162 K (i.e., about -111 ºC) (4), and then we calculated surface heat flow for a range of basal temperatures between 165 and 190 K (-108 and -83 ºC) to identify those that are consistent with the predicted surface heat flows. The heat equation has been solved through the Simscale cloud-based platform (https://www.simscale.com/), which allows complex simulations without the need for large computer resources, and has a friendly graphical user interface. We show the surface heat flow ($F_s$) obtained from the basal temperature range in Fig. 2.

Basal temperatures that are compatible with the expected range of surface heat flow increase when the porosity of the layer is taken account. However, previous work (14) has studied the effect of porosity on the temperature at the base of the ice layer and shows that an extremely porous layer is required to obtain surface heat flows less than 30 mW m$^{-2}$ with basal temperatures higher than 180 K. Such a porous layer is not compatible with the high density of the SPLD, which points to a compact ice layer with low porosity. Thus, the effect of porosity in our result will therefore be small and our conclusions remain valid.

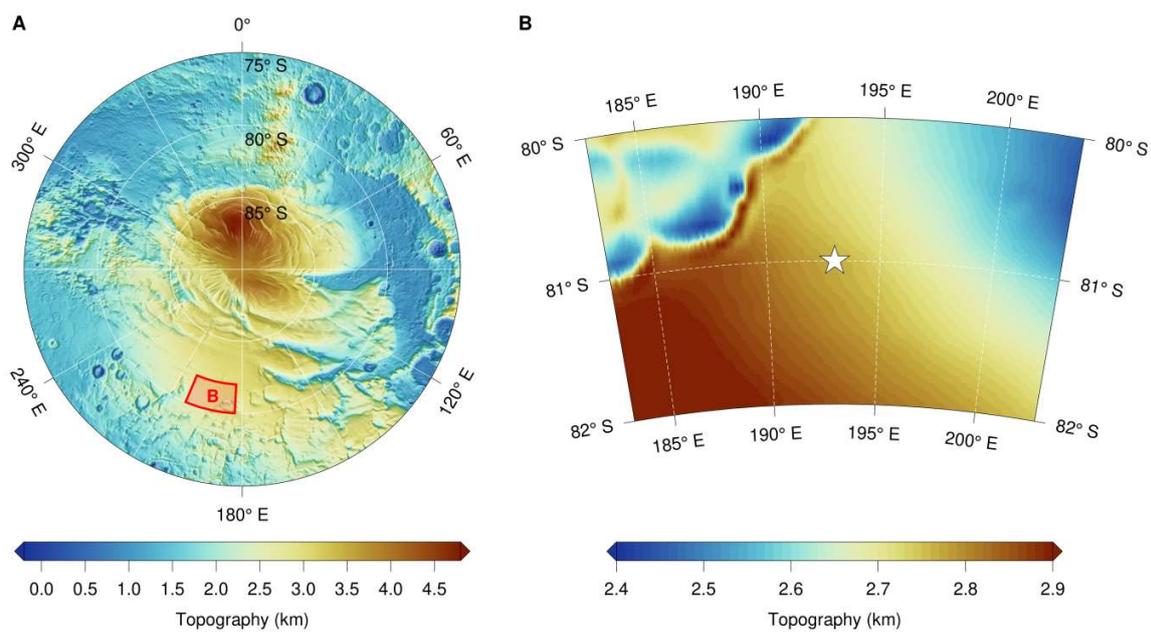

**Figure 1. Location of the putative liquid water.** Mars Orbiter Laser Altimeter (MOLA) topographic maps of the area considered in the current study. (https://astrogeology.usgs.gov/search/map/Mars/GlobalSurveyor/MOLA/Mars_MGS_MOLA_ClrShade_merge_global_463m). **A**, Polar stereographic projection of the topography from latitude 75° S to the South Pole, with a topography shaded relief superimposed for context. **B**, Zoom on the region under study, presented in a Lambert conic conformal projection with a central meridian of 193° E longitude. The white star indicates the location of the proposed melting of ice at the base of the SPLD, centred at 193° E and 81° S.

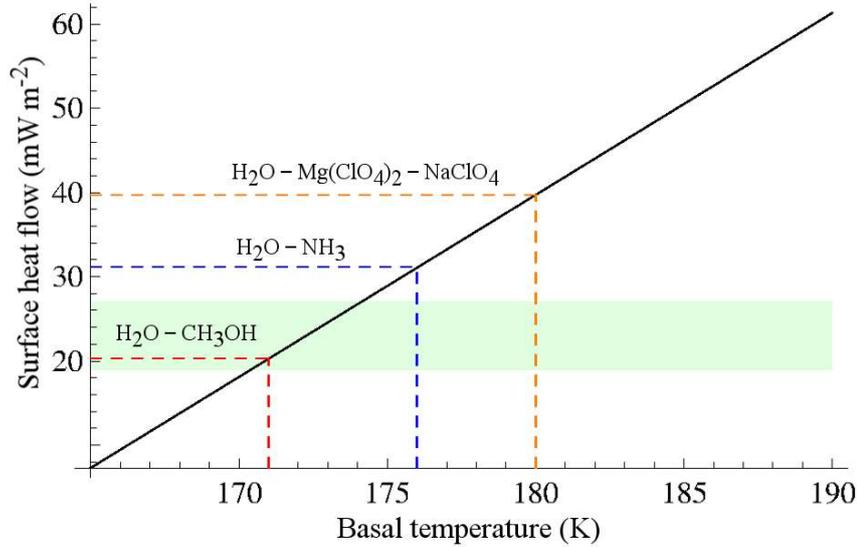

**Figure 2. Surface heat flow versus basal temperature at the location of the putative liquid water.** Surface heat flow obtained after solving the heat equation for the SPLD ice layer and considering a range of temperatures at the depth of the putative liquid water between 165 and 190 K (black line). The green shaded area indicates the surface heat flow interval calculated from the elastic thickness of the lithosphere ($T_e$). The dashed lines correspond to the freezing point temperature of an ammonia─water ice mixture (blue), methanol─water ice mixture (red), and for a mixture of sodium perchlorate and magnesium perchlorate in water (orange) (12).

| $T_e$ estimate | Surface heat flow (mW m$^{-2}$) |
|---|---|
| $T_e$ =150 km (lower limit; ref. 15 and 36) | 24.9 to 27.4 |
| $T_e$ = 255 km (best-fit in ref.18) | 20.4 to 21.7 |
| $T_e$ = 360 km (best-fit in ref.15) | 18.7 to 19.6 |

**Table 1: Surface heat flow obtained from the effective elastic thickness of the lithosphere.** Surface heat flows calculated for the South Polar region from estimates of the effective elastic thickness of the lithosphere ($T_e$). Lower and upper values of the heat flow range were calculated for thermal conductivities of 2 and 2.5 W m$^{-1}$ K$^{-1}$, corresponding respectively to a basaltic crust and to a crust including a substantial felsic component. See the Methods section for details.

| Antifreeze systems | $T_{solidus}$ | Solute concentration |
|---|---|---|
| $H_2O - NH_3$ | 176 K (-97 ºC) | <33 % by mass (47) |
| $H_2O - CH_3OH$ | 171 K (-102 ºC) | <54 % mole (48) |
| $H_2O - NaClO_4$ | 241 K (-32 ºC) | >52.28 % by mass (12) |
| $H_2O - Mg((ClO_4)_2)$ | 198 K (-75 ºC) | >45.86 % by mass (12) |
| $H_2O - Ca((ClO_4)_2)$ | 199 K (-74 ºC) | >50 % by mass (4) |
| $H_2O - NaClO_4 - Mg((ClO_4)_2)$ | 180 K (-93 ºC) | >3 mol kg$^{-1}$ Mg((ClO_4)_2) in a 3 mol kg$^{-1}$ NaClO_4 solution (12) |

**Table 2. Solidus temperature for candidates to reduce the SPLD ice layer freezing point.** Solidus temperature and required concentrations for water─ammonia and water─methanol systems together with binary and ternary solutions formed by the dissolution in water of perchlorates of sodium, magnesium and calcium.

**Acknowledgments**

**Funding:**
NASA Astrobiology Program(CPM)
Project PID2022-140686NB-I00 from the Spanish Ministry of Science and Universities (IEG, JR, AJD)

**Author contributions:**
Conceptualization: CPM, JR
Methodology: IEG, JR, AJD
Investigation: IEG, JR, CPM, JEH
Visualization: IEG, JR, JEH, CPM, AJD
Supervision: JEH, JR, CPM, AJD
Writing—original draft: IEG, JR
Writing—review & editing: JEH, JR, CPM, AJD

**Competing interests:**
The authors declare that they have no competing interests.

**Data and materials availability:**
All data needed to evaluate the conclusions in the paper are present in the paper. Heat equation solutions for different basal temperatures will be available on Zenodo repository.